\documentclass{article}

\usepackage{microtype}
\usepackage{graphicx}
\usepackage{subfigure}
\usepackage{booktabs} % for professional tables

\usepackage{amsfonts}
\usepackage{amsmath}
\usepackage{tabularx}
\usepackage{caption}

\usepackage{xcolor}

\usepackage{hyperref}

\graphicspath {{figures/}}

\usepackage[accepted]{icml2020}

\icmltitlerunning{Audio-to-Image Cross-Modal Generation}

\begin{document}

\twocolumn[
\icmltitle{Audio-to-Image Cross-Modal Generation}



\icmlsetsymbol{equal}{*}

\begin{icmlauthorlist}
\icmlauthor{Maciej {\.Z}elaszczyk}{wut}
\icmlauthor{Jacek Ma{\'n}dziuk}{wut}
\end{icmlauthorlist}

\icmlaffiliation{wut}{Faculty of Mathematics and Information Science, Warsaw University of Technology, Warsaw, Poland}

\icmlcorrespondingauthor{Maciej {\.Z}elaszczyk}{m.zelaszczyk@mini.pw.edu.pl}
\icmlcorrespondingauthor{Jacek Ma{\'n}dziuk}{mandziuk@mini.pw.edu.pl}

\icmlkeywords{Machine Learning, ICML}

\vskip 0.3in
]



\printAffiliationsAndNotice{}  % leave blank if no need to mention equal contribution

\begin{abstract}
Cross-modal representation learning allows to integrate information from different modalities into one representation. At the same time, research on generative models tends to focus on the visual domain with less emphasis on other domains, such as audio or text, potentially missing the benefits of shared representations. Studies successfully linking more than one modality in the generative setting are rare. In this context, we verify the possibility to train variational autoencoders (VAEs) to reconstruct image archetypes from audio data. Specifically, we consider VAEs in an adversarial training framework in order to ensure more variability in the generated data and find that there is a trade-off between the consistency and diversity of the generated images - this trade-off can be governed by scaling the reconstruction loss up or down, respectively. Our results further suggest that even in the case when the generated images are relatively inconsistent (diverse), features that are critical for proper image classification are preserved.
\end{abstract}

\section{Introduction}
\label{submission}

In representation learning, a frequent goal is to find a mapping from the data space (e.g. a space of images) to a more compact latent space, which is not directly observable (an embedding space / feature space). One of the main assumptions underlying such a process is that real data can be reliably summarized by a relatively limited set of factors. These factors correspond to elements of the latent / embedding / feature vectors. The resulting reduction of dimensionality facilitates the use of latent vectors for a range of tasks, such as classification or compression. Latent vectors are also frequently used in generative models, which often utilize random latent vectors in the process of data generation.

Regardless of the actual application, most of representation learning research is focused on data from one modality (e.g. image, text, audio, etc.). For instance, image classification or generation tasks usually use data solely from the image modality. Despite the success of such applications, it seems that limiting oneself to data from one modality is not a realistic model for deep learning. Real-world objects usually generate data from more than one modality and humans perceive the world around them through more than just one sense. A significant portion of the objects we interact with can be identified based on data from multiple senses.  In particular, it is usually possible to imagine an object based on only one source of sensory information (e.g. imagining a car based on the sound of its engine), which suggests the possibility of learning a common representation of an object integrating information from multiple modalities. Such a representation could then plausibly be matched to data from just one modality and the missing modalities could be inferred.

The process of learning such common representations would be facilitated by the temporal alignment of data from multiple modalities. In such a setting, sensory information from different modalities appears together and this link between multiple modalities enables learning without additional supervision.
In this work, we aim to highlight the possibility of learning shared representations from cross-modal (multi-modal) data due to the temporal alignment of datasets.

\textbf{Our contributions:}
\begin{itemize}
    \item We construct audio-image datasets and show that it is possible to generate images based on audio data, as audio features overlap with visual features to an extent that makes image generation possible.
%, at least for our synthetic datasets.
    \item We discuss how the data alignment process influences the generation process. Many-to-one mappings result in data \emph{archetypes}, while one-to-one mappings ensure more variability in the data.
    \item We analyze how scaling the reconstruction loss up or down influences the properties of the generated data.
    \item We quantitatively assess the degree to which transition between audio and image data preserves relevant visual features.
\end{itemize}

\section{Related work}

This work is related to several strands of research, which include representation learning, joint latent space learning, input reconstruction and cross-modal learning.

Representation learning is addressed by \citeauthor{Hinton1986} (\citeyear{Hinton1986}) who discuss the possibility of representing a concept as an activity pattern in more than one computing element within a network of such elements - a \emph{distributed representation}. They argue that such representations could potentially achieve content-addressable memory and generalization capabilities. \citeauthor{Schmidhuber1992} (\citeyear{Schmidhuber1992}) proposes that statistical independence of parts of the representation should be a goal of unsupervised learning. \citeauthor{Bengio2013} (\citeyear{Bengio2013}) restate this in the form of \emph{disentangled} factors of variation, motivate the importance of representation learning, and provide a review of success stories involving representation learning. They explicitly hypothesize that representations might be partially shared for different tasks or domains and that this might be beneficial in \emph{transfer learning}, \emph{domain adaptation} or \emph{multi-task learning}.

In reference to cross-modal learning and joint latent space learning, \citeauthor{Weston2010} (\citeyear{Weston2010}) train a method to learn a joint embedding space for both images and their annotations. \citeauthor{Arandjelovic2017} (\citeyear{Arandjelovic2017}) and (\citeyear{Arandjelovic2018}) show that it is possible to jointly learn audio and visual representations from an aligned audio-visual dataset with no additional supervision. The representations from the audio and visual modalities can then be compared to localize objects potentially linked to the sound and vice versa. \citeauthor{Xing2019} (\citeyear{Xing2019}) combine information from visual and semantic feature spaces to improve model performance in a few-shot learning setting. \citeauthor{Li2019} (\citeyear{Li2019}) present the existence of adversarial examples for cross-modal text/image data and show that adversarial training can improve the robustness of a cross-modal hashing network. \citeauthor{Wen2019} (\citeyear{Wen2019}) present an architecture capable of learning a shared representation for voices and faces by mapping them to their common covariates. \citeauthor{Lu2019} (\citeyear{Lu2019}) learn task-agnostic joint representations of images and natural language. In a video setting, \citeauthor{Sun2019} (\citeyear{Sun2019}) combine video data and text produced by speech-recognition methods to learned high-level \emph{linguistic features}.

In the context of input reconstruction, \citeauthor{Hinton1985} (\citeyear{Hinton1985}) pose and encoder problem where input should be reconstructed in spite of a sparse representation, while \citeauthor{Rumelhart1986} (\citeyear{Rumelhart1986}) show that it is possible to train a neural network with an intermediate bottleneck layer in an unsupervised manner to reconstruct input, an idea further developed in \citep{Bourlard1988}, \citep{Kramer1991}, \citep{Hinton1993} and \citep{Vincent2008}. \citeauthor{Kingma2014} (\citeyear{Kingma2014}) use a mapping from the input to the parameters of a multidimensional normal distribution in a variational autoencoder (VAE) architecture. This facilitates the process of generating new data from latent space representations without the corresponding input data. Extensions incorporating discrete latent codes \citep{vandenOord2017} have also been proposed, resulting in images of increasing size \citep{Razavi2019}.

A different line of generative models has been triggered by \citeauthor{Goodfellow2014} (\citeyear{Goodfellow2014}), who proposed a generative adversarial network (GAN) framework. This framework relies on game-theoretic ideas and jointly trains two neural networks with opposing objectives in order for one of the networks to learn to generate data similar but not identical to examples from the train dataset. The GAN architecture has been extended, for instance to learn a mapping from data space to the latent space \cite{Donahue2017} or handle images of higher resolution \cite{Brock2019}. While GANs tend to produce sharper images than VAEs, significant instabilities in adversarial training have been identified and various approaches to stabilize training have been proposed \cite{Radford2016}, \cite{Arjovsky2017}, \cite{Miyato2018}, with no approach conclusively resolving the issue so far.

There is also a growing body of research on cross-modal generative models. \citeauthor{vandenOord2016} (\citeyear{vandenOord2016}) train a generative model conditioned on \emph{linguistic features} extracted from text input for the text-to-speech task. Conditional GANs are used to generate images from text in \cite{Reed2016}. The tasks of generating images from audio and generating audio from images are tackled in an adversarial setting in \cite{Chen2017}. \citeauthor{Hao2018} (\citeyear{Hao2018}) extend the adversarial framework to generate audio and visual data mutually. \citeauthor{Hsu2018} (\citeyear{Hsu2018}) train a modified VAE to incorporate not only shared factors but also modality-dependent ones for audio/image datasets. \citeauthor{Shi2019} (\citeyear{Shi2019}) employ a cross-modal VAE to generate images from text and text from images. \citeauthor{MullerEberstein2019} (\citeyear{MullerEberstein2019}) show that, with pre-training, VAEs can be recombined to generate audio from images and that the produced audio is at least somewhat consistent with human notions of similarity between the audio and images, even in the absence of aligned datasets.

While research on cross-modal generative models has advanced significantly, it is still very much underdeveloped. Most of this area focuses on linking the linguistic domain with the audio domain (e.g. text to speech) or providing a bridge between the linguistic and the visual domains (e.g. image captions, text-to-image generation). At present, the audio-visual branch is underrepresented within the broader cross-modal generative current. In spite of the development of cross-modal methods, relatively little research on audio-visual architectures has been published. Even less research has been conducted directly on audio-to-image generation. Relative to image datasets, audio-visual datasets are few and far between.

In this context
%, it is our aim to bridge the existing research gap in audio-to-image generation. In particular, 
we strive to align audio and image datasets in order to facilitate the process of audio-to-image generation. We also propose methods to extract audio features and utilize their subset, the \emph{audio-visual features}, to produce images consistent with the ones observed in the data space. These methods differ from the standard supervised conditional variational autoencoder or conditional generative adversarial network setups \cite{Goodfellow2014}. Our work is a potential starting point for research on the possibility of utilizing variational autoencoders in an adversarial setting for audio-visual generation. It is also the first study we are aware of concerned with many-to-one vs. one-to-one mappings for cross-modal dataset alignment and the effects of these mappings on the generated data. Our work points to a new potential area of research related to precise methods of set alignment and how they can impact the data generation process. In particular, many-to-one mappings may be preferable for some tasks, and vice versa.

\begin{figure}[t]
\includegraphics[width=\columnwidth]{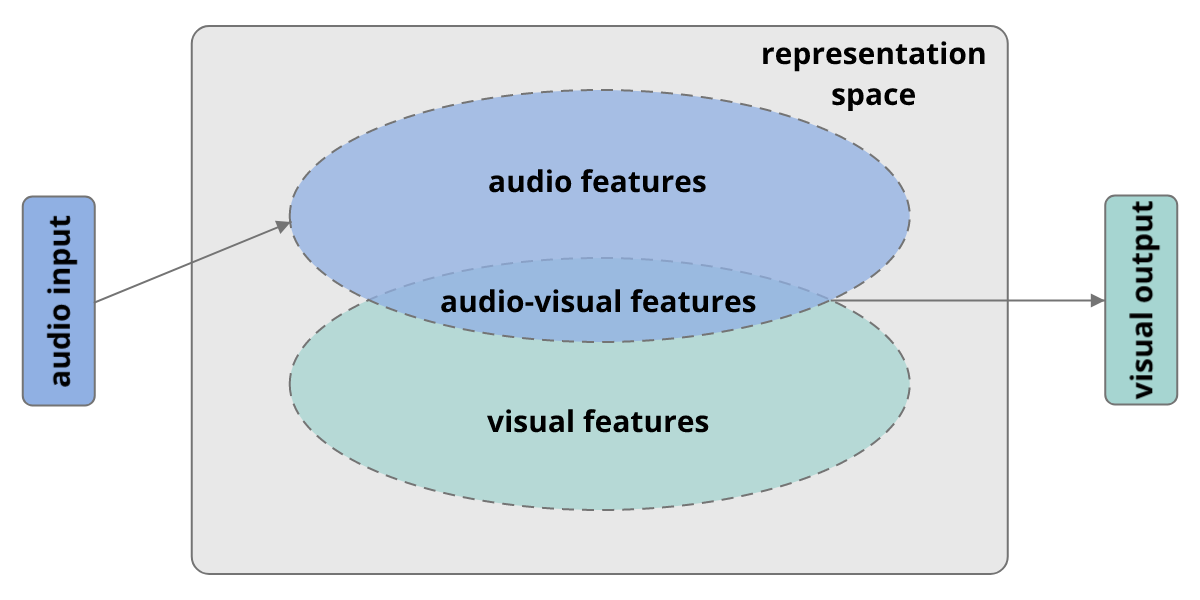}
\caption{Relationship between audio and visual features.}
\label{fig:features}
\end{figure}

\section{Generative models}

For two variables $\mathbf{x}, \mathbf{y}$, a \emph{generative model} is one which describes the joint probability distribution $p(\mathbf{x},\mathbf{y})$. In machine learning applications, the model can be written as $p(\mathbf{x},\mathbf{z})$, where $\mathbf{x}$ comes from the \emph{data space}, e.g. the space where we can observe the data, and $\mathbf{z}$ comes from the \emph{latent space}, a directly unobservable space of features.

\subsection{Variational autoencoders (VAEs)}

VAEs \cite{Kingma2014} extend the autonecoder architecture. Let us consider the generative model $p_{\theta}(\mathbf{z})p_{\theta}(\mathbf{x} \vert \mathbf{z})$, and $q_{\phi}(\mathbf{z} \vert \mathbf{x})$ be the approximation of the true intractable posterior $p_{\theta}(\mathbf{z} \vert \mathbf{x})$. We may refer to $q_{\phi}(\mathbf{z} \vert \mathbf{x})$ as a probabilistic \emph{encoder} and to $p_{\theta}(\mathbf{x} \vert \mathbf{z})$ as a probabilistic \emph{decoder}. Let $p_{\theta}(\mathbf{x} \vert \mathbf{z})$ and $q_{\phi}(\mathbf{z} \vert \mathbf{x})$ be neural networks parametrized by $\theta$ and $\phi$, respectively. Let $\mathbf{z}$ be a random variable and $\mathbf{z} = g_{\theta}(\mathbf{x}, \epsilon)$, where $\epsilon$ is an auxiliary variable with an independent marginal and $g_{\theta}$ is a function parametrized by $\theta$. We can sample from the posterior $\mathbf{z}^{(i)} \sim q_{\phi}(\mathbf{z} \vert \mathbf{x}^{(i)})$, calculating $\mathbf{z}^{(i)} = g_{\theta}(\mathbf{x}^{(i)}, \epsilon) = \mu^{(i)} + \sigma^{(i)} \odot \epsilon$, where $\epsilon \sim \mathcal{N}(\mathbf{0}, \mathbf{I})$. The loss function for a given data example can be expressed as:
\begin{multline}
    \mathcal{L}(\mathbf{x}^{(i)}, \theta, \phi) =
    -\mathbb{E}_{\mathbf{z} \sim q_{\phi}(\mathbf{z} \vert \mathbf{x^{(i)}})}\left[\log{p_{\theta}(\mathbf{x}^{(i)} \vert \mathbf{z})}\right] \\
    + \mathbb{KL}(q_{\phi}(\mathbf{z} \vert \mathbf{x^{(i)}}) \vert\vert p(\mathbf{z}))
\end{multline}

where $\mathbb{KL}$ is the Kullback–Leibler divergence.

In practice, the process of training a VAE consists of passing data through the encoder, which maps to parameter vectors $\mu$ and $\sigma$. These vectors are then combined with a random $\epsilon$ vector sampled from the standard normal distribution via the equation $\mathbf{z}^{(i)} = \mu^{(i)} + \sigma^{(i)} \odot \epsilon$. Thanks to the reparametrization trick, the model is differentiable and can be trained with backpropagation. It is worth noting that the loss formulation consists of two terms: the \emph{reconstruction error} and the KL divergence. While the reconstruction loss encourages adhering to examples seen in the dataset, the KL divergence ensures that the encoder remains consistent with the standard normal distribution assumed for $p(\mathbf{z})$. Such an assumption simplifies sampling from a trained model.

\subsection{Generative adversarial networks (GANs)}

GANs, first introduced in \cite{Goodfellow2014}, apply the concept of a minimax game to generative models. In this setting, we consider two neural networks, the \emph{generator} $G(\mathbf{z})$ and the \emph{discriminator} $D(\mathbf{x})$. We define a latent space prior $p_{\mathbf{z}}(\mathbf{z})$. $G$ maps from the latent space to the data space, while $D$ maps from the data space to a scalar output, which represents the probability that a sample comes from the data space rather than that it has been produced by $G$. $G$ and $D$ are trained in turns, essentially playing a two-player minimax game:
\begin{multline}
    \min_{G}\max_{D}V(D,G)=\mathbb{E}_{\mathbf{x} \sim p_{data}(\mathbf{x})}\left[\log{(D(\mathbf{x}))}\right] \\
    +\mathbb{E}_{\mathbf{z} \sim p_{\mathbf{z}}(\mathbf{z})}\left[\log{(1-D(G(\mathbf{z})))}\right]
\end{multline}

This minimax game has a global optimum for $p_g = p_{data}$, i.e. for $G$ generating images indistinguishable from the ones in the data space and $D$ continually producing ${1 \over 2}$ as its assessment of the probability that the samples come from the data distribution. This global optimum can theoretically be achieved via backpropagation in sequential weight updates for both networks. In reality, the training process for GANs is frequently unstable and convergence is by no means automatic.

\section{Audio-to-Image Architectures}

\begin{figure*}[t]
\includegraphics[width=2.1\columnwidth]{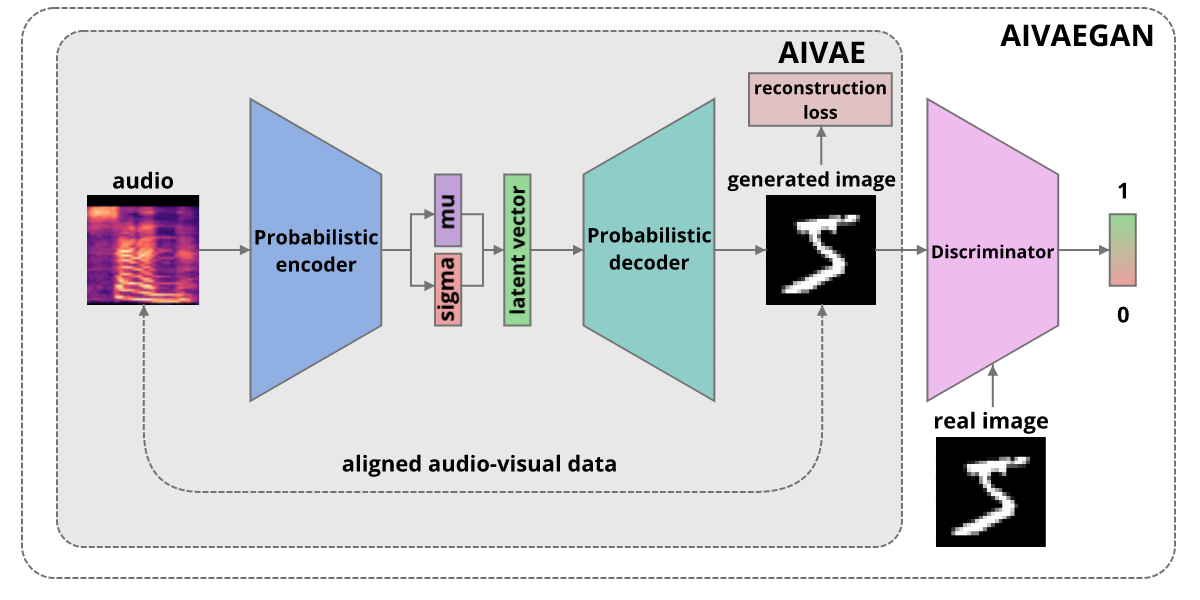}
\caption{AIVAE and AIVAEGAN architectures.}
\label{fig:architectures}
\end{figure*}

We consider two generative setups geared towards extracting audio representations from sound and using a subset of these features relevant for the visual domain, the \emph{audio-visual features}, for image generation (Figure \ref{fig:features}). Our first model is a modification of the classic VAE architecture. Our second architecture is a combination of a VAE and a GAN network. We use VAE as the GAN generator and train the whole system adversarially. Both architectures are presented in Figure \ref{fig:architectures}.

\subsection{Audio-to-Image VAE}

We modify the VAE encoder to process audio data and map it to the parameters of the latent space. It now describes a distribution $q_{\phi}(\mathbf{f} \vert \mathbf{a})$, where $\mathbf{a}$ represents the audio input and $\mathbf{f}$ denotes the \emph{audio-visual features} - features from the audio space, which would be relevant for image generation. These features are used as parameters $\mu$ and $\sigma$ in the process of generating random vectors $\mathbf{f}^{(i)} = \mu^{(i)} + \sigma^{(i)} \odot \epsilon$, in line with the classic VAE architecture. The random vectors are then passed through the decoder $p_{\theta}(\mathbf{i} \vert \mathbf{f})$, which maps to the image space. The loss minimized in training is:

\begin{multline}
    \mathcal{L}(\mathbf{a}^{(i)}, \mathbf{i}^{(i)}, \theta, \phi) =
    -\mathbb{E}_{\mathbf{f} \sim q_{\phi}(\mathbf{f} \vert \mathbf{a^{(i)}})}\left[\log{p_{\theta}(\mathbf{i}^{(i)} \vert \mathbf{f})}\right] \\
    + \mathbb{KL}(q_{\phi}(\mathbf{f} \vert \mathbf{a^{(i)}}) \vert\vert p(\mathbf{f}))
\end{multline}

The whole architecture is trained with backpropagation just like a vanilla VAE. However, the Audio-to-Image VAE (AIVAE) requires two aligned data sources, as each training example is a tuple $\left( \mathbf{a}^{(i)}, \mathbf{i}^{(i)} \right)$.

\subsection{Audio-to-Image VAE-GAN}

AIVAE can be further extended in a setting where it is treated as the generator in an adversarial framework, $G(\mathbf{a}) = \text{AIVAE}(\mathbf{a})$. Now, $G: \mathbf{a} \to \mathbf{i}$, where $\mathbf{a}$ and $\mathbf{i}$ are the audio and image spaces, respectively. The discriminator $D(\mathbf{i})$ tries to discern between real images and the ones manufactured by $G$. It is worth noting that $D$ does not have explicit access to the pairing between the image and audio and so is not likely to enforce correct classes from $G$. The information about the pairing $\left( \mathbf{a}^{(i)}, \mathbf{i}^{(i)} \right)$ is incorporated via a reconstruction loss term in the objective function:
\begin{multline}
    \min_{G}\max_{D}V(D,G) = \mathbb{E}_{\mathbf{i} \sim p_{data}(\mathbf{i})}\{\log{(D(\mathbf{i}))}\}\\
    + \mathbb{E}_{\mathbf{a} \sim p_{data}(\mathbf{a})}\mathbb{E}_{\mathbf{f} \sim q_{\phi}(\mathbf{f} \vert \mathbf{a})} \{ \log{(1-D(G(\mathbf{f})))} \\
    - \alpha \log{p_{\theta}(\mathbf{i} \vert \mathbf{f})} + \mathbb{KL}(q_{\phi}(\mathbf{f} \vert \mathbf{a}) \vert\vert p(\mathbf{f}))\}
\end{multline}

This objective function incorporates the opposing goals of $G$ and $D$, while taking into account the information about the expected reconstruction and desired latent (feature) distribution. The hyperparameter $\alpha$ is designed to regulate the importance of reconstruction vs. adversarial goals. Similarly to the AIVAE setup, the Audio-to-Image VAE-GAN (AIVAEGAN) loss relies on aligned data without explicit supervision.

\section{Experiments}

We test AIVAE and AIVAEGAN on two synthetic datasets in order to assess their capability to generate images from audio data.

\subsection{Datasets}

We construct two synthetic datasets for our experiments. Both of them rely on the MNIST dataset as a source of digit images to be paired with audio data.

\subsubsection{MNIST-FSDD}

The first dataset combines $28 \times 28$-pixel MNIST digits with audio data from the Free Spoken Digit Dataset (FSDD)~\footnote{The FSDD dataset is available at \\ \url{https://github.com/Jakobovski/free-spoken-digit-dataset}.}. FSDD consists of $2\,000$ \texttt{WAV} files, which are the recorded pronunciations of digits by $4$ speakers, which amounts to $50$ pronunciations per digit per speaker. We represent the recordings as MEL-scaled spectrograms resized to $48 \times 48$ pixels. We then perform a random train/test split, where $90\%$ of the data is assigned to the train set and the remaining $10\%$ of the data is in the test set.

After initial preprocessing, the FSDD data is paired with the MNIST digits in the following procedure. We sample digits from MNIST without replacement and for each of them, we sample with replacement from FSDD spectrograms with matching labels. Each MNIST digit is used once, but each FSDD spectrogram is paired multiple times with various MNIST digits from the corresponding class. This many-to-one MNIST $\to$ FSDD mapping enforces the alignment of datasets along labels. This is the only time the label information is used prior to evaluation. The outlined steps are repeated separately for the train and test sets. The synthesized dataset consists of $60\,000$ audio-image pairs for the train set and $10\,000$ pairs for the test set.

\subsubsection{MNIST-SCD}

Our second dataset is based on MNIST for image data and on the Speech Commands Dataset (SCD) for audio data~\footnote{The SCD dataset can be downloaded from \\ \url{https://ai.googleblog.com/2017/08/launching-speech-commands-dataset.html}.}. SCD has $65\,000$ one-second long recordings of $30$ words. We limit our analysis to the words representing digits. This leaves us with $23\,666$ utterances distributed among $10$ classes. For simplicity, we will henceforth refer to this subset as \emph{SCD}, having in mind that we consider the digit-related subset of the original dataset. Similarly to the previous case, we convert the \texttt{WAV} files to MEL-scaled spectrograms, resize them to $48 \times 48$ pixels and randomly split the datset so that $90\%$ of the data is used as a train set and $10\%$ is held out for evaluation. We align MNIST and the truncated SCD in a sequential approach. For each class label we randomly sample a spectrogram without replacement and match it to a MNIST image with the same class via random sampling without replacement. This results in a one-to-one MNIST $\to$ SCD mapping. Again, these steps are performed separately for the train and test sets. The resulting train set is comprised of $21\,160$ audio-image pairs, while the produced test set consists of $2\,360$ audio-image pairs.

\subsection{Results}

We test the AIVAE (Table~\ref{tab:aivae-architecture}) and AIVAEGAN (Table~\ref{tab:aivaegan-architecture}) architectures on the MNIST-FSDD and MNIST-SCD datasets with the latent space dimension set to $64$ in each experiment. All models are trained for $100$ epochs with batches of $128$ examples and are optimized with Adam \cite{Kingma2015}. The initial learning rate is set to $10^{-3}$ for AIVAE and to $2 \times 10^{-4}$ for both the generator and the discriminator of AIVAEGAN.

\begin{table}[t]
\caption{Audio-to-image variational autoencoder architecture (AIVAE).}
\label{tab:aivae-architecture}
\vskip 0.15in
\begin{center}
\begin{small}
\begin{sc}
\begin{tabularx}{\columnwidth}{ll}
\toprule
Audio encoder & Image decoder \\
\midrule
Input 48x48 & Input 64 \\
Conv 4x4, 64, str=2 & FC 512 \\
ReLU & ReLU \\
Conv 4x4, 128, str=2 & FC 1024 \\
ReLU & ReLU \\
FC 1024 & FC 7x7x128 \\
ReLU & ReLU \\
FC 512 & Upconv 4x4, 64, str=2 \\
ReLU & ReLU \\
$\mu, \sigma$: FC 64 & Upconv 4x4, 1, str=2 \\
ReLU & Sigmoid \\
Output 2x64 & Output 28x28 \\
\bottomrule
\end{tabularx}
\end{sc}
\end{small}
\end{center}
\vskip -0.1in
\end{table}

\begin{table}[t]
\caption{Audio-to-image variational autoencoder - generative adversarial network architecture (AIVAEGAN).}
\label{tab:aivaegan-architecture}
\vskip 0.15in
\begin{center}
\begin{scriptsize}
\begin{sc}
\begin{tabularx}{\columnwidth}{lll}
\toprule
Audio encoder & Image decoder & Discriminator \\
\midrule
Input 48x48 & Input 64 & Input 28x28 \\
Conv 4, 128, 2 & Upconv 3, 512, 2 & Conv 4, 128, 2 \\
Conv 4, 256, 2 & Upconv 3, 256, 2 & Conv 4, 256, 2 \\
Conv 4, 512, 2 & Upconv 2, 128, 2 & Conv 4, 512, 2 \\
$\mu, \sigma$: Conv 4, 64, 2 & Upconv 2, 1, 2 & Conv 1, 1, 1, Sigm\\
Output 2x64 & Output 28x28 & Output 1 \\
\midrule
LeakyReLU($0.2$) & ReLU & LeakyReLU($0.2$) \\
\midrule
BatchNorm & BatchNorm & BatchNorm \\
\bottomrule
\end{tabularx}
\end{sc}
\end{scriptsize}
\end{center}
\vskip -0.1in
\end{table}

\subsubsection{Qualitative evaluation}

We first investigate the reconstruction capabilities of AIVAE for MNIST-FSDD and MNIST-SCD. Figure~\ref{fig:vae-toy-big} presents real images from the last epoch of training together with their reconstructions. The classes of the generated images are consistent with their real counterparts for both datasets. The images generated for MNIST-SCD display more variability and less blur, while the ones generated for MNIST-FSDD display a pattern in which each class is represented by visually indistinguishable images - the \emph{archetypes} for each class. The archetypical nature of these images stems from the fact that the MNIST-FSDD dataset was constructed via a many-to-one mapping and each audio data point was associated with multiple digits from the MNIST database. Consequently, for each digit the model has learned its \emph{average} reconstruction.

\begin{figure}[t]
\subfigure[MNIST-FSDD]{\includegraphics[width=\columnwidth]{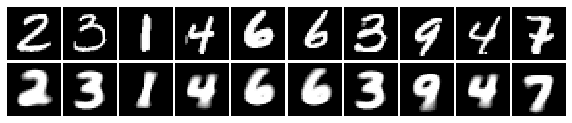}}
\subfigure[MNIST-SCD]{\includegraphics[width=\columnwidth]{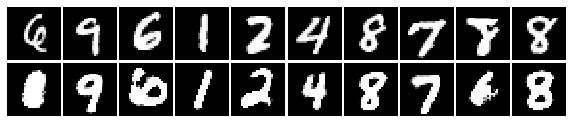}}
\caption{Test-set reconstructions for AIVAE. Random, not cherry-picked.}
\label{fig:test-vae}
\end{figure}

The reconstructions are further evaluated on the test set with the results presented in Figure~\ref{fig:test-vae}. These outcomes support the initial idea that the model trained on the many-to-one dataset recovers digit archetypes. The model trained on the one-to-one dataset is less consistent, as evidenced by partially unintelligible reconstructions, however, it does retain reconstruction capabilities and shows more variability in the generated data.

For the AIVAEGAN model, the train-set reconstructions on MNIST-FSDD and MNIST-SCD are presented in Figures~\ref{fig:vae-gan-toy} and~\ref{fig:vae-gan-big}, respectively. For MNIST-FSDD, the quality of reconstructions at the end of training is relatively high, with lower levels of $\alpha$ promoting more variation in the generated images. Conversely, for $\alpha = 1$ and $\alpha = 2$, stricter adherence to train data is observed. The results for MNIST-SCD follow the same pattern, however, the one-to-one mapping in this dataset ensures more variability is preserved without necessarily sacrificing the correct labels.

These outcomes are verified on the test sets (Figure \ref{fig:test-vae-gan}) and are consistent with our earlier remarks. Lower $\alpha$ levels result in more variation and less adherence to proper classes. Higher values of $\alpha$ result in more archetypical data, however, even for larger $\alpha$ values a one-to-one training set mapping preserves diversity in the generated data.

\begingroup
\setlength{\tabcolsep}{0.1pt}
\begin{table}[t]
  \centering
  \begin{tabular}{cc}
    generated & real \\
    \raisebox{-.5\height}{\includegraphics[width=0.48\columnwidth]{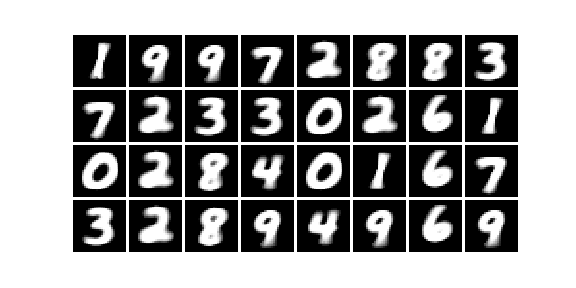}}
      & \raisebox{-.5\height}{\includegraphics[width=0.48\columnwidth]{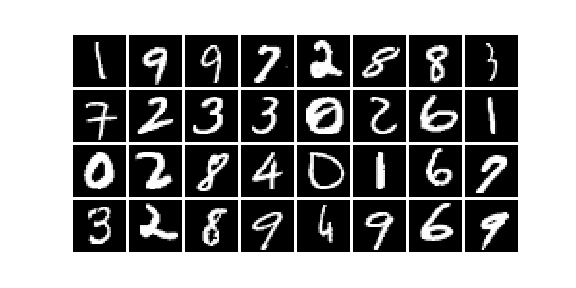}} \\ \\
    \raisebox{-.5\height}{\includegraphics[width=0.48\columnwidth]{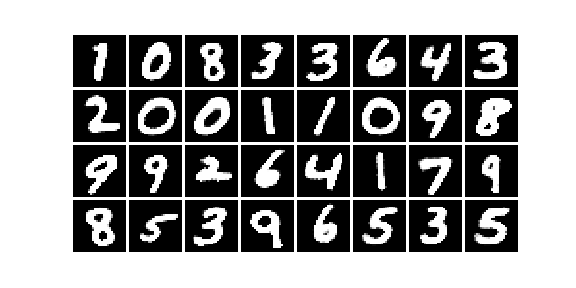}}
      & \raisebox{-.5\height}{\includegraphics[width=0.48\columnwidth]{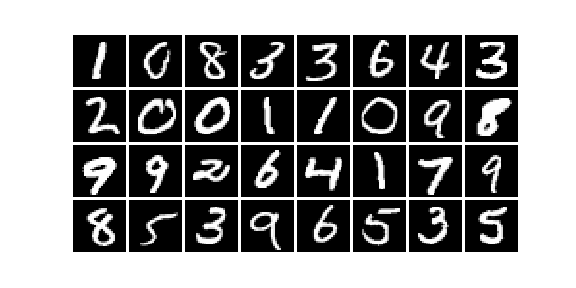}}
  \end{tabular}
  \captionof{figure}{Generated vs. real images at the end of training (AIVAE). Top row: MNIST-FSDD. Bottom row: MNIST-SCD.} \label{fig:vae-toy-big}
\end{table}
\endgroup

\begingroup
\setlength{\tabcolsep}{0.1pt}
\begin{table}[t]
  \centering
  \begin{tabular}{ccc}
    $\alpha$ & generated & real \\
    $0.2$ & \raisebox{-.5\height}{\includegraphics[width=0.48\columnwidth]{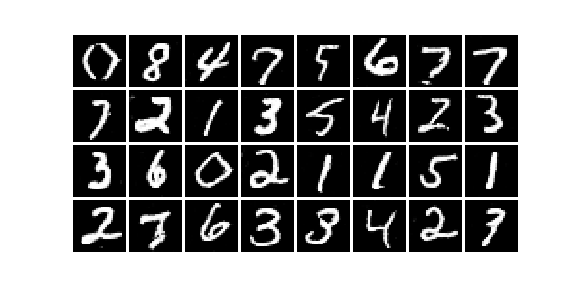}}
      & \raisebox{-.5\height}{\includegraphics[width=0.48\columnwidth]{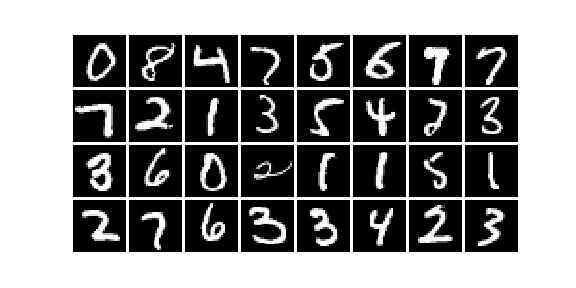}} \\ \\
    $0.5$ & \raisebox{-.5\height}{\includegraphics[width=0.48\columnwidth]{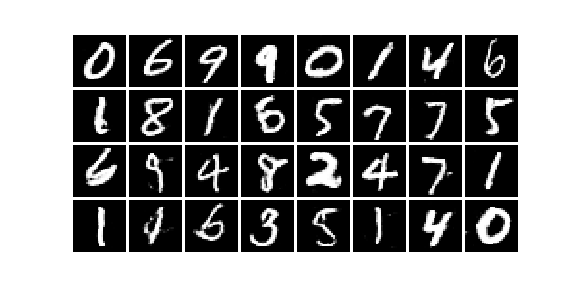}}
      & \raisebox{-.5\height}{\includegraphics[width=0.48\columnwidth]{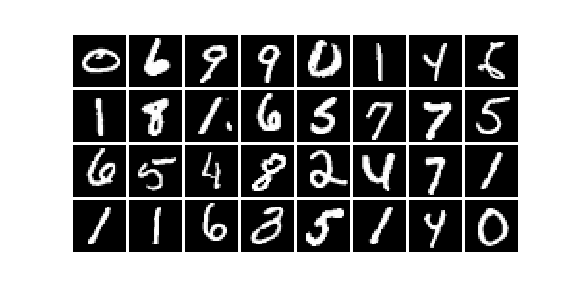}} \\ \\
    $1$ & \raisebox{-.5\height}{\includegraphics[width=0.48\columnwidth]{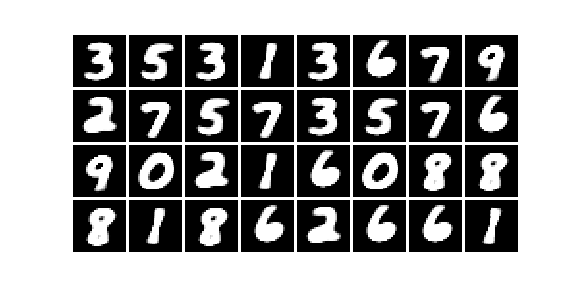}}
      & \raisebox{-.5\height}{\includegraphics[width=0.48\columnwidth]{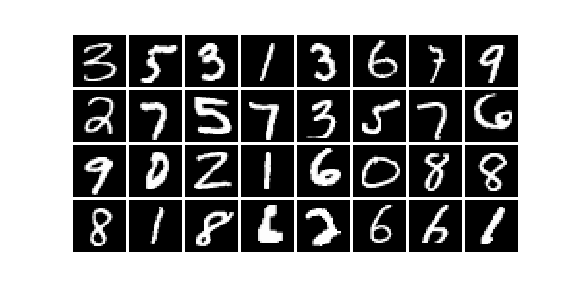}} \\ \\
    $2$ & \raisebox{-.5\height}{\includegraphics[width=0.48\columnwidth]{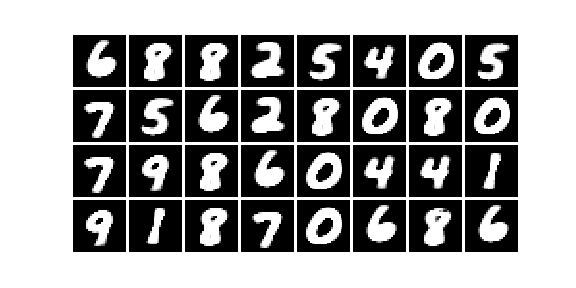}}
      & \raisebox{-.5\height}{\includegraphics[width=0.48\columnwidth]{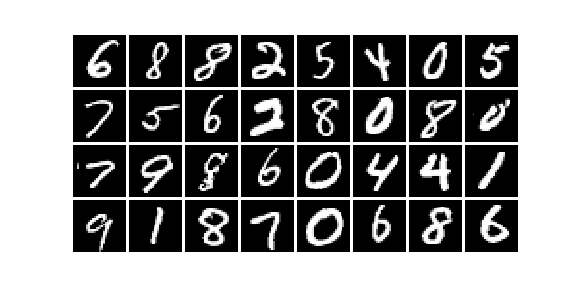}}
  \end{tabular}
  \captionof{figure}{Generated vs. real images at the end of training for different levels of reconstruction loss importance (AIVAEGAN, MNIST-FSDD).} \label{fig:vae-gan-toy}
\end{table}
\endgroup

\begingroup
\setlength{\tabcolsep}{0.1pt}
\begin{table}[t]
  \centering
  \begin{tabular}{ccc}
    $\alpha$ & generated & real \\
    $0.2$ & \raisebox{-.5\height}{\includegraphics[width=0.48\columnwidth]{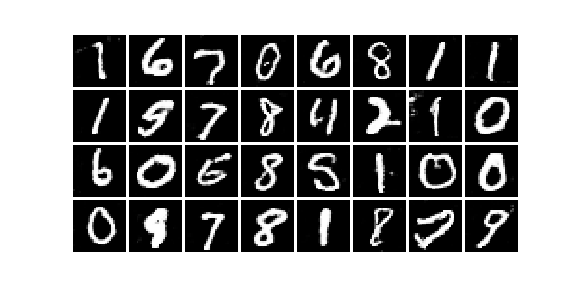}}
      & \raisebox{-.5\height}{\includegraphics[width=0.48\columnwidth]{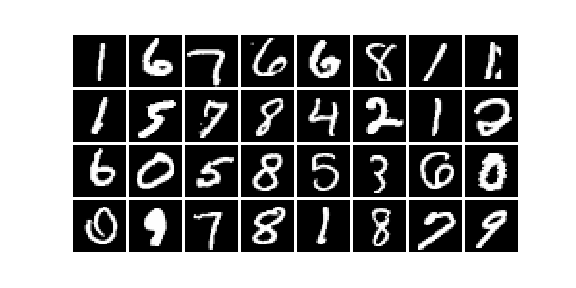}} \\ \\
    $0.5$ & \raisebox{-.5\height}{\includegraphics[width=0.48\columnwidth]{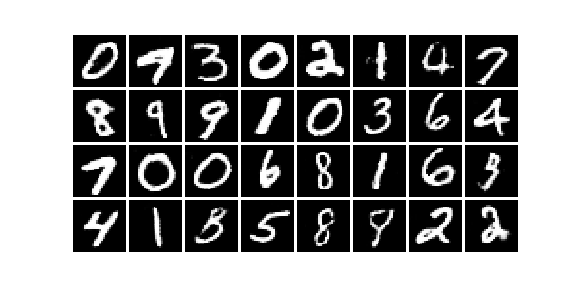}}
      & \raisebox{-.5\height}{\includegraphics[width=0.48\columnwidth]{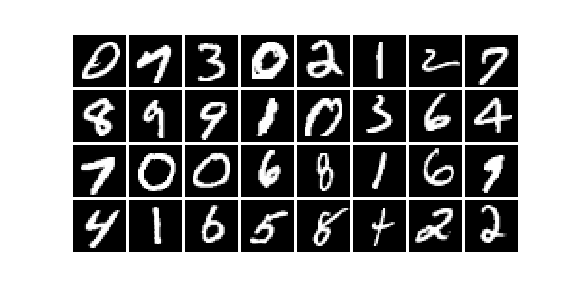}} \\ \\
    $1$ & \raisebox{-.5\height}{\includegraphics[width=0.48\columnwidth]{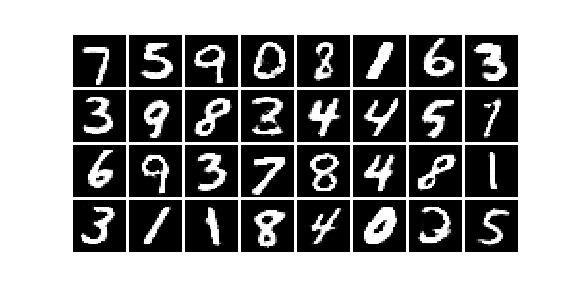}}
      & \raisebox{-.5\height}{\includegraphics[width=0.48\columnwidth]{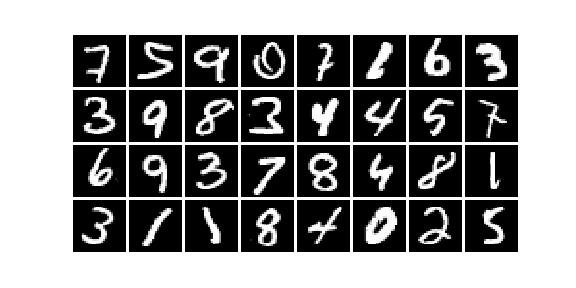}} \\ \\
    $2$ & \raisebox{-.5\height}{\includegraphics[width=0.48\columnwidth]{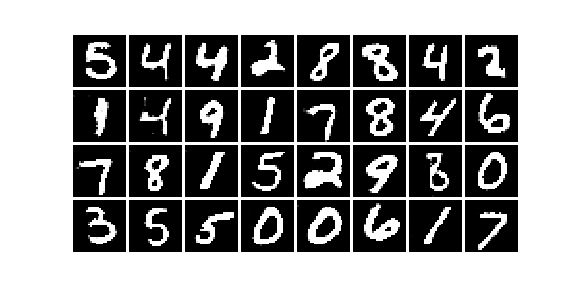}}
      & \raisebox{-.5\height}{\includegraphics[width=0.48\columnwidth]{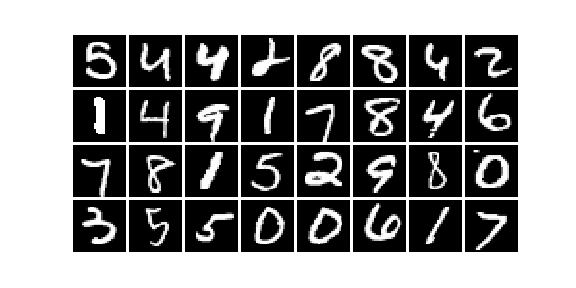}}
  \end{tabular}
  \captionof{figure}{Generated vs. real images at the end of training for different levels of reconstruction loss importance (AIVAEGAN, MNIST-SCD).} \label{fig:vae-gan-big}
\end{table}
\endgroup

\begingroup
\setlength{\tabcolsep}{0.1pt}
\begin{table}[t]
  \centering
  \begin{tabular}{ccc}
    $\alpha$ & MNIST-FSDD & MNIST-SCD \\
    $0.2$ & \raisebox{-.5\height}{\includegraphics[width=0.48\columnwidth]{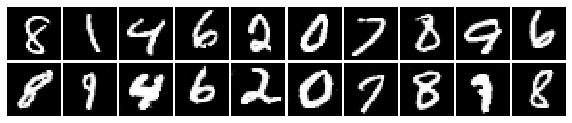}}
      & \raisebox{-.5\height}{\includegraphics[width=0.48\columnwidth]{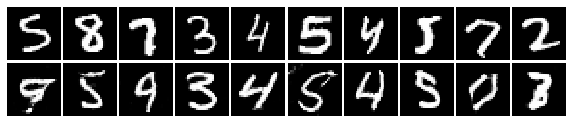}} \\ \\
    $0.5$ & \raisebox{-.5\height}{\includegraphics[width=0.48\columnwidth]{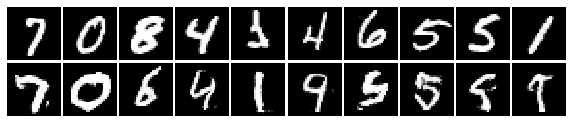}}
      & \raisebox{-.5\height}{\includegraphics[width=0.48\columnwidth]{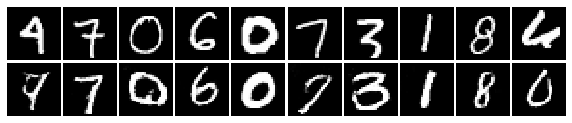}} \\ \\
    $1$ & \raisebox{-.5\height}{\includegraphics[width=0.48\columnwidth]{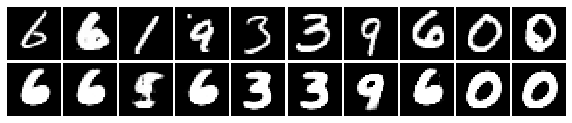}}
      & \raisebox{-.5\height}{\includegraphics[width=0.48\columnwidth]{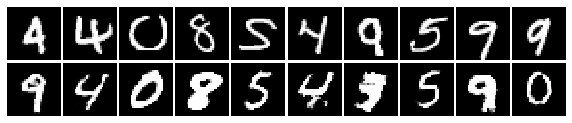}} \\ \\
    $2$ & \raisebox{-.5\height}{\includegraphics[width=0.48\columnwidth]{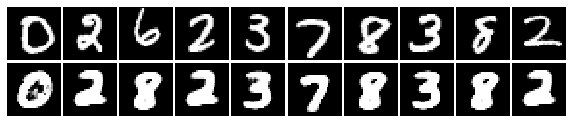}}
      & \raisebox{-.5\height}{\includegraphics[width=0.48\columnwidth]{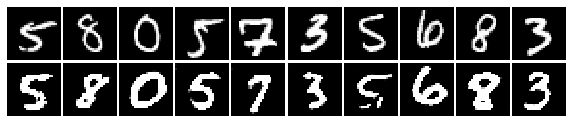}}
  \end{tabular}
  \captionof{figure}{Test-set reconstructions by AIVAEGAN for different levels of reconstruction loss importance. Random, not cherry-picked.} \label{fig:test-vae-gan}
\end{table}
\endgroup

\subsubsection{Quantitative evaluation}

While eyeball tests confirm the possibility of image generation from audio features for aligned datasets, such a form of inspection is inherently subjective. A quantitative evaluation of the ability of AIVAE and AIVAEGAN to preserve audio features important for image generation would be a more objective measure. To this end, the following procedure is proposed. First, we generate images from audio for the test sets of both MNIST-FSDD and MNIST-SCD. Then, these images are fed to a pretrained MNIST classifier and the accuracy of prediction is recorded. Such a process would allow to assess the extent to which the images generated by our models retain information necessary for image classification.

We choose a LeNet5 classifier \cite{Lecun1998} - a classic CNN with convolutional/pooling blocks followed by fully-connected layers. The details of the architecture are presented in Table \ref{tab:lenet5}. We train it on the MNIST train set for $100$ epochs with Adam, with an initial learning rate of $10^{-3}$. The network achieves a $98.7\%$ accuracy on the MNIST test set. This is the benchmark we would compare our models against. Intuitively, the accuracy of the pretrained classifier on the data generated by our models should be lower as the audio features used in the data generation process potentially encompass only a subset of the relevant visual features (the \emph{audio-visual features}).

\begin{table}[t]
\caption{LeNet5 architecture.}
\label{tab:lenet5}
\vskip 0.15in
\begin{center}
\begin{small}
\begin{sc}
\begin{tabularx}{\columnwidth}{l}
\toprule
 LeNet5 \\
\midrule
Input 32x32\\
Conv 1x1, 6, str=5, ReLU \\
Max Pool 2x2, str=2 \\
Conv 6x6, 16, str=5, ReLU \\
Max Pool 2x2, str=2 \\
Conv 16x16, 120, str=5, ReLU \\
Flatten 120 \\
FC 84, ReLU \\
FC 10, Softmax \\
Output 10 \\
\bottomrule
\end{tabularx}
\end{sc}
\end{small}
\end{center}
\vskip -0.1in
\end{table}

\begin{table}[t]
\caption{Test-set classification accuracy of a pretrained LeNet5 network on images generated by AIVAE.}
\label{tab:acc-aivae}
\vskip 0.15in
\begin{center}
\begin{small}
\begin{sc}
\begin{tabularx}{\columnwidth}{lcc}
\toprule
 Data set & MNIST-FSDD & MNIST-SCD \\
\midrule
Accuracy & $94.2\%$ & $86.6\%$ \\
\bottomrule
\end{tabularx}
\end{sc}
\end{small}
\end{center}
\vskip -0.1in
\end{table}

\begin{table}[t]
\caption{Test-set classification accuracy for a pretrained LeNet5 network on images generated by AIVAEGAN.}
\label{tab:acc-aivaegan}
\vskip 0.15in
\begin{center}
\begin{small}
\begin{sc}
\begin{tabularx}{\columnwidth}{lcccc}
\toprule
Data set \textbackslash $\alpha = $ & $0.2$ & $0.5$ & $1$ & $2$ \\
\midrule
MNIST-FSDD & $81.0\%$ & $80.5\%$ & $93.0\%$ & $94.3\%$ \\
MNIST-SCD & $66.9\%$ & $76.2\%$ & $81.7\%$ & $81.5\%$ \\
\bottomrule
\end{tabularx}
\end{sc}
\end{small}
\end{center}
\vskip -0.1in
\end{table}

It turns out that for AIVAE the accuracy remains close to $95\%$ for the many-to-one data setting. The one-to-one data alignment results in lower accuracy, albeit still above $85\%$ (see Table~\ref{tab:acc-aivae}). AIVAEGAN results display a pattern where for many-to-one data the accuracy of the LeNet5 classifier remains considerably higher than for the one-to-one dataset. At the same time, a higher weight associated with the reconstruction loss improves visual feature retention.

A comparison of AIVAE and AIVAEGAN reveals that the gap in performance between the two models is influenced by the level of $\alpha$ set for AIVAEGAN. Higher levels of $\alpha$ result in a smaller difference in performance between the two models, particularly for $\alpha = 2$ and the MNIST-FSDD dataset. This is perhaps not surprising, as the reconstruction loss explicitly enforces stronger adherence to images from the data space. A lower $\alpha$ promotes more diversity in the generated digits, but this comes at a cost of losing visual features distinguishable for a pretrained classifier. It does seem that the effect of the reconstruction loss on the error rate is more pronounced for the many-to-one mapping. Going from $\alpha = 0.2$ to $\alpha = 2$, the error rate drops $44\%$ in the one-to-one setting, while the corresponding drop for the many-to-one setting is $70\%$. The lower sensitivity of the error rate to changes in $\alpha$ for the one-to-one setting, combined with the less stable adversarial learning process, may also account for a lower top accuracy on MNIST-SCD when comparing AIVAEGAN against AIVAE.

\begin{figure}[h]
    \centering
    \includegraphics[width=0.49\columnwidth]{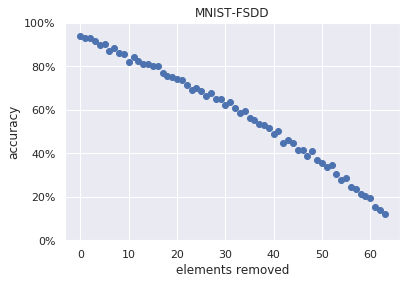}
    \includegraphics[width=0.49\columnwidth]{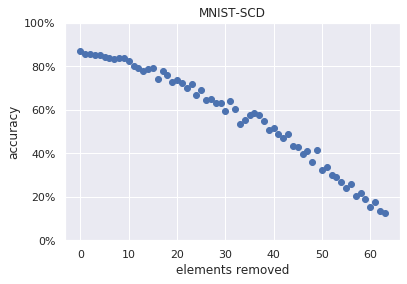}
    \caption{Classification accuracy of a pretrained LeNet5 classifier for AIVAE with different numbers of the elements of the latent vector randomly set to $0$.}
    \label{fig:restricted}
\end{figure}

We conduct an additional evaluation of the quality of features of AIVAE trained on MNIST-FSDD and MNIST-SCD. Namely, we randomly set $k$ elements of the latent vector to $0$ and record the classification accuracy of the LeNet5 classifier on both test sets. We repeat this for $k \in \{0, 1, \dots 64\}$. The results are presented in Figure \ref{fig:restricted}. With $k$ increasing, the classification accuracy naturally diminishes, confirming the usefulness of the \emph{audio-visual features} for image generation. The interesting result is that the gap in accuracy between the models trained on the many-to-one dataset and the one-to-one dataset narrows down and actually disappears for larger numbers of the elements of the latent vector restricted to $0$. This suggests that for the many-to-one alignment, the whole set of features is crucial for performance, while the one-to-one alignment results in a model that is more robust to a removal of a small subset of features.

\section{Conclusions}

In this work we show that it is possible to generate images based on audio features extracted from sounds. We train cross-modal VAEs and their adversarial extensions on two synthetic datasets and obtain results which suggest that the \emph{audio-visual features} - the subset of relevant visual features corresponding to audio features - are sufficient to generate images similar to the ones from the data space. We also show that the degree of adherence of the generated images to the ones from the data space can be modeled by choosing an appropriate weight for the reconstruction loss. For audio which repeats itself for a given image, our models tend to produce image archetypes. In a setting where each data point is uniquely paired with another one from the other modality, archetypes are weaker and the data retains more diversity without necessarily losing much of the relevant visual features.

Our study provides results in a hitherto underrepresented area of cross-modal learning related to audio-visual data. We show that audio-to-image generation is possible both in a VAE architecture and by adversarially extending it. This provides a new approach toward architectures which could be employed in a cross-modal setting. Additionally, we provide an analysis of the impact the kind of dataset alignment (many-to-one vs. one-to-many) may have on the generated data. It is not hard to imagine scenarios where the generation of archetypes might be of interest (e.g. when we are interested in only the general notion and not a specific object) vs. scenarios where diversity could be preferable to strict adherence (e.g. when we are interested in obtaining many examples of one specific notion).

%From a research point of view, 
On a more general note, our work provides a method for generating data in an unsupervised manner, which could contribute toward creating a more realistic model of the world in artificial intelligence systems.
%Our results may also be of practical importance. 
Generating images from sound 
%is interesting if we consider that it 
could potentially be used in a variety of situations, e.g. to help people with impaired hearing, to visualize sounds, to create art, etc.

Presented results point to several potential lines of research which we plan to consider in our future work.
%First, it could be beneficial to perform a comparable study on a larger scale, particularly on more challenging audio-image datasets.
First of all, while in this study we have discounted the possibility of forming one representation per modality rather than one universal shared representation, it might be beneficial to analyze settings where two or more representations are formed with a shared subset as this may allow to retain more relevant features from each analyzed modality.
Second of all, the obtained results suggest that the supervisory signal contained in the alignment of datasets may be enough to at least somewhat successfully generate data from one modality based on the aligned data from another modality. This, in turn, suggests the possibility of leveraging a number of unlabeled datasets to obtain shared cross-modal representations. Online streaming services which present their users with videos seem like a natural choice since videos contain naturally aligned audio and visual data. It would be interesting to see extensions handling temporal data such as audio-video streams.
Finally, it would be interesting to test the extent to which cross-modal representations can be formed from unaligned datasets, and the extent to which such representations could be used to bind two data points from different modalities.
%Finally, cross-modal representation could potentially be useful as a means of \emph{filling in the blanks} in either the data or the internal representations of models. It would be %interesting to see more research geared toward enhancing the usefulness of cross-modal data in the process of building an internal world model.

\iffalse
\section*{Acknowledgements}

\textbf{Do not} include acknowledgements in the initial version of
the paper submitted for blind review.

If a paper is accepted, the final camera-ready version can (and
probably should) include acknowledgements. In this case, please
place such acknowledgements in an unnumbered section at the
end of the paper. Typically, this will include thanks to reviewers
who gave useful comments, to colleagues who contributed to the ideas,
and to funding agencies and corporate sponsors that provided financial
support.

\nocite{langley00}
\fi

\bibliography{icml2020}
\bibliographystyle{icml2020}

\end{document}